\documentclass[12pt]{article}
\usepackage[latin9]{inputenc}
\usepackage{amsmath}
\usepackage[unicode=true,pdfusetitle,
 bookmarks=true,bookmarksnumbered=false,bookmarksopen=false,
 breaklinks=false,pdfborder={0 0 1},backref=false,colorlinks=false]
 {hyperref}

\makeatletter
\@ifundefined{date}{}{\date{}}

\usepackage{latexsym}
\newcommand{\be}{\begin{equation}}
\newcommand{\ee}{\end{equation}}
\newcommand{\bea}{\begin{eqnarray}}
\newcommand{\eea}{\end{eqnarray}}
\newcommand{\beas}{\begin{eqnarray*}}
\newcommand{\eeas}{\end{eqnarray*}}

\def\XXint#1#2#3{{\setbox0=\hbox{$#1{#2#3}{\int}$ }
\vcenter{\hbox{$#2#3$ }}\kern-.5\wd0}}

\makeatother

\begin{document}

\title{Large N bilocals at the infrared fixed point of the three dimensional
O(N) invariant vector theory with a quartic interaction}

\author{Mbavhalelo Mulokwe\thanks{Email: mbavhalelo.mulokwe@wits.ac.za}
\enskip{}and Jo\~ao  P. Rodrigues\thanks{Email: joao.rodrigues@wits.ac.za}\\
 \\
 National Institute for Theoretical Physics \\
 School of Physics and Mandelstam Institute for Theoretical Physics
\\
 University of the Witwatersrand, Johannesburg\\
 Wits 2050, South Africa \\
 }
\maketitle
\begin{abstract}
We study the three dimensional O(N) invariant bosonic vector model
with a $\frac{\lambda}{N}(\phi^{a}\phi^{a})^{2}$ interaction at its
infrared fixed point, using a bilocal field approach and in an $1/N$
expansion. We identify a (negative energy squared) bound state in
its spectrum about the large $N$ conformal background. At the critical
point this is identified with the $\Delta=2$ state. We further demonstrate
that at the critical point the $\Delta=1$ state disappears from the
spectrum. \pagebreak{} 
\end{abstract}

\section{Introduction}

The AdS/CFT correspondence \cite{Maldacena:1997re, Gubser:1998bc, Witten:1998qj}
finds a very interesting area of application in the context of the
higher spin theories/vector model correspondence \cite{Klebanov:2002ja}. Of particular interest to us, in this context,
is the $AdS_{4}/CFT_{3}$ correspondence.\footnote{There is a vast literature on the subject;  \cite{Fronsdal:1978rb,Fradkin:1986qy,Giombi:2009wh,Giombi:2010vg} are representative of the work on the subject, but they do not form by any means an exhaustive list.}
Although the higher spin degrees of freedom of Fronsdal and Vasiliev
are not those of string theory\footnote{ For attempts to link the two, see for instance  \cite{Chang:2012kt,Giombi:2011kc,Sezgin:2012ag,Honda:2017nku}},
there are several reasons why this correspondence is of great interest.
These include the absence of supersymmetry and the fact that vector
models are \char`\"{}solvable\char`\"{} in the large $N$ limit, allowing
for a more concrete and detailed study of the workings of the correspondence,
and possibly even providing a definition of (gauge fixed) higher spin
theories themselves, through their dual vector valued field theories.

We are in particular interested in and motivated by the constructive
approach of \cite{Das:2003vw,Koch:2010cy,deMelloKoch:2012vc,Jevicki:2011ss}.
In this approach, the singlet sector of $O(N)$ invariant field theories
is described in terms of equal time bilocals:

\begin{equation}
\psi_{\vec{x_{1}}\vec{x_{2}}}=\sum_{a=1}^{N}\phi^{a}\left(t,\vec{x_{1}}\right)\phi^{a}\left(t,\vec{x_{2}}\right).
\end{equation}

\noindent where $\vec{x_1}$ and $\vec{x_2}$ are two dimensional space
vectors. These 5 degrees of freedom and their canonical conjugates
are mapped, in the free UV fixed point, to $AdS_{4}\times S_{1}$
where the $S_{1}$ encodes the spin degrees of freedom. Originally
formulated in a light cone gauge, the map has since been obtained
in a temporal gauge \cite{Koch:2010cy}. In momentum space it is a
point transformation and is given by:

\begin{flalign}
E & =E_{1}+E_{2}=|\vec{p_{1}}|+\vec{p_{2}}|\label{free energies}\\
\vec{p} & =\vec{p_{1}}+\vec{p}_{2}\\
p^{z} & =2\sqrt{\left|\vec{p_{1}}\right|\left|\vec{p_{2}}\right|}\sin\left(\frac{\varphi_{2}-\varphi_{1}}{2}\right)\\
\theta & =\arctan\left(\frac{2\vec{p}_{2}\times\vec{p}_{1}}{\left(\left|\vec{p_{1}}\right|-\left|\vec{p_{2}}\right|\right)p^{z}}\right)
\end{flalign}
with $\varphi_{2}-\varphi_{1}$ being the angle between $\vec{p_{1}}$
and $\vec{p_{2}}$ \cite{Koch:2010cy}.

\noindent
The holographic coordinate is given by 

\begin{equation}
z={(\vec{x_1}-\vec{x_2})\cdot (\vec{p_1} |p_2| - \vec{p_2} |p_1| )\over p^z (|\vec{p_1}| + |\vec{p_2}|)},  \label{z coord}
\end{equation}

\noindent
The three dimensional $O(N)$ vector theory with a $\frac{\lambda}{N}(\phi^{a}\phi^{a})^{2}$
interaction has a IR fixed critical point. At this critical point,
the theory is expected to contain a state with dimension $\Delta=2$,
a boundary field in the standard AdS/CFT correspondence with the standard
positive branch for the expression of the dimension of the operator
\cite{Klebanov:2002ja}, and no longer the $\Delta=1$ state of present
in the UV critical point. Although general arguments exist relating
the two through a Legendre transformation \cite{Petkou:2003zz} ,
in practice the IR fixed point is described in terms of a non-linear
sigma model\cite{ZinnJustin,Lang:1990ni}. In this description, the
Lagrange multiplier field is naturally identified with the $\Delta=2$
state, but it is certainly not apparent that the $\Delta=1$ is no longer present
in the theory, or equivalently, that the constraint is enforced beyond
the leading large N order. 


It is the purpose of this article to elucidate these issues directly
in terms of the $\frac{\lambda}{N}(\phi^{a}\phi^{a})^{2}$ theory,
using a bilocal approach that allows for a systematic expansion in
$1/N$. Keeping in mind that the $AdS_{4}/CFT_{3}$ constructive approach
developed for the UV free fixed point is canonical (ensuring that
the correct number of degrees of freedom are matched), it is important
to have a single time bilocal field description of the field theory
IR fixed point. This is developed in this article. However, it is
not the purpose of this article to discuss the map \cite{Koch:2010cy} at the IR fixed
point. This is left for a later communication.

This paper is organized as follows: in chapter 2, the collective field
theory \cite{Jevicki:1979mb} Hamiltonian, expressed in terms of equal
time bilocal fields and their canonical conjugates, is presented.
The large N conformal background is obtained. The bilocal fluctuations
about this background are obtained for large but finite $\lambda$
and the equations of motion cast in the form of a highly non-trivial
integral equation in momentum space. In Chapters 3 and 4, we consider
the path integral description of the theory. This requires the introduction
of (two-time) covariant bilocals. In Chapter 3, we consider the description
of the non-linear sigma model in terms of (two-time) bilocal fields
plus the dynamical Lagrange multiplier field. We obtain the two point
function for the (shifted) bilocals and for the dynamical Lagrange
multiplier field. The two point function for the bilocal fields takes
the same form as that of the free theory, and the two point function
of the dynamical Lagrange multiplier is that of a $\Delta=2$ conformal
field. In Chapter 4, we present the (two time) bilocal description
of the $\frac{\lambda}{N}(\phi^{a}\phi^{a})^{2}$ theory. The two
point function of the bilocal fields, which is equivalent to the Bethe-Salpeter
equation for the underlying fundamental vector fields, consists of
a disconnected (free) piece and a connected diagram describing the
s-channel scattering of a composite field. For finite $\lambda$,
it has a pole at\footnote{Our Minkowski signature is (+,-,-))}

\begin{equation}
E^{2}-\left(\vec{p}_{1}+\vec{p}_{2}\right)^{2}=-\frac{\lambda^{2}}{48^{2}}.\label{bound state energy}
\end{equation}

At the critical ($\lambda\to\infty$) point, the connected diagram
is identical, up to external leg factors, to the two point function
of the dynamical Lagrange multiplier field, and hence is identified
with the $\Delta=2$ state. The disconnected piece is the same as
that of the free case.

In Chapter 5, we successfully integrate over an intermediate energy
variable in the bound state integral equation identified in the connected
diagram of the two point function of Chapter 4, and establish that
the result is the same as that obtained in the Hamiltonian approach
of Chapter 2. It is left then to understand the role of the disconnected
(free) piece of the (two time) bilocal propagator and how the corresponding
states are solutions of the quadratic Hamiltonian equations of motion.
After all, we require bilocals to construct the bulk. The answer,
which we discuss in Chapter 6, is that these are scattering states
of positive squared energy. The problem can then be formulated as
that of potential scattering off a delta function potential in the
relative coordinates. An equation for the local $\Delta=1$ composite
is obtained, and is shown to vanish at the critical point as $\lambda\to\infty$.
Further evidence of the disappearance of this state from the spectrum
is provided at the level of the path integral approach, where
we show that the correlator of two $\Delta=1$ composites vanishes
at the critical point. This makes explicit the conjectured cancellation
between disconnected and connected diagrams in the original proposal
of Klebanov and Polyakov \cite{Klebanov:2002ja}.

In summary, we explicitly demonstrate in this article that in a conformal
background, the large $N$ spectrum of the three dimensional $O(N)$
vector model with a $\frac{\lambda}{N}(\phi^{a}\phi^{a})^{2}$ interaction
consists of a (negative energy squared) bound state that at the IR
critical point becomes a $\Delta=2$ state and of scattering states,
with an energy dispersion the same as that of the free theory, but
with a $\Delta=1$ state that is removed from the spectrum at the
IR critical point.

\section{Hamiltonian }

\label{hamiltonian}

Our main interest is to investigate the collective large-$N$ spectrum
of the critical $O(N)$ vector-model

\begin{equation}
\mathcal{L}=\frac{1}{2}\left(\partial_{\mu}\phi^{a}\partial^{\mu}\phi^{a}-m^{2}\phi^{a}\phi^{a}\right)-\frac{\lambda}{4!N}(\phi^{a}\phi^{a})^{2}.\label{lagrangian}
\end{equation}

\noindent As such, the starting point is an Hamiltonian expressed
in terms of equal time collective bilocals and their canonical conjugates.
The equal time bilocals have been defined earlier:

\begin{equation}
\psi_{\vec{x}\vec{y}}=\sum_{a=1}^{N}\phi^{a}\left(t,\vec{x}\right)\phi^{a}\left(t,\vec{y}\right).
\end{equation}

\noindent Changing variables from the $\phi^{a}$$\left(a=1,\cdots,N\right)$
fields to the bilocals introduces a non-trivial Jacobian that, to
leading order in $N$, is given by

\begin{equation}
\log J=\frac{N}{2}\mathrm{Tr}\log\psi.
\end{equation}

\noindent The trace is in (spatial) functional space.

It is by now well known that the collective field theory hamiltonian
can be written as \cite{Jevicki:1983hb,Koch:2010cy}($d=3$ in this
article)\footnote{The fields have been rescaled $\psi\to N\psi$ to evidence explicitly
the $N$ dependence.}:

\begin{equation}
H=\frac{2}{N}\mathrm{Tr}\left(\Pi\psi\Pi\right)+\frac{N}{8}\mathrm{Tr}\psi^{-1}+N\int d^{d-1}\vec{x}\left(\frac{1}{2}m^{2}\psi_{\vec{x}\vec{x}}+\frac{1}{2}\lim_{\vec{y}\rightarrow\vec{x}}-\partial^{2}\psi_{\vec{y}\vec{x}}+\frac{\lambda}{4!}\psi_{\vec{x}\vec{x}}^{2}\right),
\end{equation}
where the conjugate momentum is

\begin{equation}
\Pi_{\vec{x}\vec{y}}=-i\frac{\delta}{\delta\psi_{\vec{x}\vec{y}}}.
\end{equation}

\noindent In the large-$N$ limit the kinetic term is subleading and,
with the large $N$ translationally invariant ansatz

\begin{equation}
\psi_{\vec{x}\vec{y}}=\int\frac{d^{d-1}\vec{k}}{\left(2\pi\right)^{d-1}}e^{i\vec{k}\left(\vec{x}-\vec{y}\right)}\psi_{\vec{k}},
\end{equation}

\noindent the saddle-point equations yields:

\begin{equation}
\psi_{\vec{k}}^{0}=\frac{1}{2}\left(\vec{k}^{2}+m^{2}+\frac{\lambda}{6}\int\frac{d^{d-1}\vec{k'}}{\left(2\pi\right)^{d-1}}\psi_{\vec{k}'}\right)^{-1/2}.
\end{equation}

\noindent Integrating both sides one obtains the standard gap equation:

\begin{equation}
s=\frac{1}{2}\int\frac{d^{d-1}\vec{k}}{\left(2\pi\right)^{d-1}}\frac{1}{\sqrt{\vec{k}^{2}+m^{2}+\frac{\lambda}{6}s}},
\end{equation}
where

\begin{equation}
s=\int\frac{d^{d-1}\vec{k'}}{\left(2\pi\right)^{d-1}}\psi_{\vec{k}'}.
\end{equation}

\noindent Defining 
\[
\alpha=m^{2}+\frac{\lambda}{6}s,
\]

\noindent one has

\[
\frac{6}{\lambda}(\alpha-m^{2})=\int\frac{d^{d-1}\vec{k}}{\left(2\pi\right)^{d-1}}\frac{1}{2\sqrt{\vec{k}^{2}+\alpha}}=\int\frac{d^{d}k}{\left(2\pi\right)^{d}}\frac{i}{k^{2}-\alpha}=\int\frac{d^{d}k_{E}}{\left(2\pi\right)^{d}}\frac{1}{k_{E}^{2}+\alpha}.
\]

\noindent Our regularization is defined as:

\begin{equation}
\int\frac{d^{d}k_{E}}{\left(2\pi\right)^{d}}\frac{1}{k_{E}^{2}+\alpha}=\frac{1}{\left(4\pi\right)^{d/2}}\Gamma\left(1-\frac{d}{2}\right)\alpha^{\frac{d-2}{2}}.\label{regularization}
\end{equation}

\noindent Thus, for $d=3$ one obtains the equation $\alpha+\frac{\lambda}{24\pi}\sqrt{\alpha}-m^{2}=0$.
The IR fixed point is associated with the root:

\begin{equation}
\sqrt{\alpha}=\frac{m^{2}}{\lambda}+O(\frac{m^{4}}{\lambda^{3}})\label{mass}
\end{equation}

\noindent and is approached by keeping $m^{2}$ finite and taking
$\lambda\to\infty$. At the critical point then, the background propagator
takes the conformal form:

\begin{equation}
\psi_{\vec{k}}^{0}=\frac{1}{2\sqrt{\vec{k}^{2}}},
\end{equation}

\noindent and is the $O(N)$ invariant two point function of the underlying
scalar fields.

For the next $1/N$ correction, we study the spectrum of fluctuations
about the large $N$ conformal background. We let

\begin{equation}
\psi=\psi^{0}+\frac{1}{\sqrt{N}}\eta,\qquad\Pi=\sqrt{N}p,
\end{equation}

\noindent and expand the Hamiltonian up to quadratic order. We find:

\begin{equation}
H^{\left(2\right)}=2\mathrm{Tr}\left(p\psi^{0}p\right)+\frac{1}{8}\mathrm{Tr}\left(\psi_{0}^{-1}\eta\psi_{0}^{-1}\eta\psi_{0}^{-1}\right)+\frac{\lambda}{4!}\int d^{d-1}\vec{x}\eta_{\vec{x}\vec{x}}^{2}.
\end{equation}

\noindent The fluctuations satisfy the Hamiltonian equations of motion.
We note\footnote{We consistently use the coordinate exchange symmetry of the bilocals}:

\begin{flalign}
\dot{\eta}_{\vec{x}\vec{y}} & =\frac{\delta H_{2}}{\delta p_{\vec{x}\vec{y}}}\nonumber \\
 & =2\left(\left(p\psi_{0}\right)_{\vec{y}\vec{x}}+\left(\psi_{0}p\right)_{\vec{y}\vec{x}}\right)=\dot{\eta}_{\vec{y}\vec{x}},
\end{flalign}

\noindent and obtain

\begin{multline}
\ddot{\eta}_{\vec{x_{1}}\vec{x_{2}}}=-\frac{1}{4}\left(\eta\psi^{0^{-1}}\psi^{0^{-1}}+\psi^{0^{-1}}\eta\psi^{0^{-1}}+\psi^{0^{-1}}\eta\psi^{0^{-1}}+\psi^{0^{-1}}\psi^{0^{-1}}\eta\right)_{\vec{x_{1}}\vec{x_{2}}}\\
+\frac{\lambda}{6}\left(\eta\delta\psi^{0}\right)_{\vec{x_{1}}\vec{x_{2}}}+\frac{\lambda}{6}\left(\psi^{0}\eta\delta\right)_{\vec{x_{1}}\vec{x_{2}}}.
\end{multline}

\noindent 
We look for eigen-frequencies in momentum space,

\begin{flalign}
\eta_{\vec{x_{1}}\vec{x_{2}}}(t) &=e^{-iEt} \eta_{\vec{x_{1}}\vec{x_{2}}} \nonumber \\
\eta_{\vec{x_{1}}\vec{x_{2}}} & =\int\frac{d^{d-1}\vec{k}_{1}}{\left(2\pi\right)^{\frac{d-1}{2}}}\int\frac{d^{d-1}\vec{k}_{2}}{\left(2\pi\right)^{\frac{d-1}{2}}}e^{i\vec{k}_{1}\vec{x}_{1}+i\vec{k}_{2}\vec{x}_{2}}\eta_{\vec{k}_{1}\vec{k}_{2}}.
\end{flalign}

\noindent
and obtain the following equation for the spectrum of fluctuations:

\begin{equation}
E^{2}\eta_{\vec{k}_{1}\vec{k}_{2}}=\frac{1}{4}\left(\psi_{\vec{k}_{1}}^{0^{-1}}+\psi_{\vec{k}_{2}}^{0^{-1}}\right)^{2}\eta_{\vec{k}_{1}\vec{k}_{2}}+\frac{\lambda}{6}\left(\psi_{\vec{k}_{1}}^{0}+\psi_{\vec{k}_{2}}^{0}\right)\int\frac{d^{d-1}\vec{l}}{\left(2\pi\right)^{d-1}}\eta_{\vec{k}_{1}+\vec{k}_{2}-\vec{l},\vec{l}},\label{scattering equation}
\end{equation}

\noindent Note that for $\lambda=0$

\begin{equation}
E_{\vec{k}_{1}\vec{k}_{2}}^{2}=\frac{1}{4}\left(\psi_{\vec{k}_{1}}^{0^{-1}}+\psi_{\vec{k}_{2}}^{0^{-1}}\right)^{2}=\left(|\vec{k_{1}}|+|\vec{k_{2}}|\right)^{2}, \label{free}
\end{equation}

\noindent a result known for some time \cite{Jevicki:1983hb} and
at the root of the $AdS_{4}/CFT_{3}$ constructive map \cite{Koch:2010cy}
in the free UV fixed point. For finite $\lambda$, equation (\ref{scattering equation})
can be recast in the form:

\begin{equation}
\eta_{\vec{k}_{1}\vec{k}_{2}}=\frac{\frac{\lambda}{6}\left(\psi_{\vec{k}_{1}}^{0}+\psi_{\vec{k}_{2}}^{0}\right)}{E^{2}-\frac{1}{4}\left(\psi_{\vec{k}_{1}}^{0^{-1}}+\psi_{\vec{k}_{2}}^{0^{-1}}\right)^{2}}\int\frac{d^{d-1}\vec{l}}{\left(2\pi\right)^{d-1}}\eta_{\vec{k}_{1}+\vec{k}_{2}-\vec{l},\vec{l}}.\label{eq:single time equation for the fluctuations}
\end{equation}

\noindent A solution only exists for this equation if the following
condition is satisfied:

\begin{equation}
1=\frac{\lambda}{12}\int\frac{d^{d-1}\vec{k}}{\left(2\pi\right)^{d-1}}\frac{1}{E_{p}^{2}-\left(\left|\vec{k}\right|+\left|\vec{p}-\vec{k}\right|\right)^{2}}\left(\frac{1}{\left|\vec{k}\right|}+\frac{1}{\left|\vec{p}-\vec{k}\right|}\right).\label{ham pole condition}
\end{equation}

\noindent This is obtained when we multiply both sides of (\ref{eq:single time equation for the fluctuations})
with $\delta\left(\vec{k}_{1}+\vec{k}_{2}-\vec{p}\right)$ and integrate
over $\vec{k}_{1}$ and $\vec{k}_{2}$.

We will be able to obtain the solution to (\ref{eq:single time equation for the fluctuations})  and (\ref{ham pole condition}), and show
that they correspond to a relativistic bound state with the energy
momentum relation given by (\ref{bound state energy}). Scattering
states with energies given by (\ref{free})  (or (\ref{free energies})), are general solutions
of (\ref{scattering equation}) which can be thought of as a relativistic
version of a quantum mechanical potential scattering problem. We will
establish that as $\lambda\to\infty$ the $\Delta=1$ field $\eta_{\vec{x}\vec{x}}$
is no longer present in the spectrum.

In order to do so, in the next two sections we first examine the two-time
bilocal formulation of the same problem in the
the path integral formalism, which is the standard approach to conformal
field theories in terms of their correlators, and where one hopes
to find a covariant description of the spectrum of the theory by the
identification of poles in the appropriate propagators. We will then
show how the equations and conditions of this section can be obtained
by successfully integrating over an appropriate energy variable.

\section{Bilocal description of the non-linear $\sigma$ model}

In this section, we consider the non-linear sigma model in the collective
field theory approach. The reason behind this is the argument \cite{ZinnJustin}
that in the large-$N$ limit, the $O\left(N\right)$ vector-model
at its infra-red critical point is described by a non-linear sigma
model  \cite{Lang:1990ni,Giombi:2009wh}.

Recall that the action for the non-linear sigma model can be written
as\footnote{In this section we use an euclidean signature}

\begin{equation}
S=N\int d^{d}x\left(\frac{1}{2}\partial_{\mu}\vec{S}\partial_{\mu}\vec{S}+\frac{\alpha\left(x\right)}{2}\left(\vec{S}^{2}-\frac{1}{\lambda}\right)\right)
\end{equation}
with the $\alpha\left(x\right)$ field playing the role of a Lagrange
multiplier enforcing the constraint $\vec{S}^{2}=\frac{1}{\lambda}$.

We introduce the covariant bilocals

\begin{equation}
\psi_{xy}=\vec{S}\left(x\right)\cdot\vec{S}\left(y\right).
\end{equation}
For the log of the Jacobian, we have (e.g. \cite{deMelloKoch:1996mj}):
\begin{equation}
\log J=\frac{N}{2}\mathrm{Tr}\log\psi.\label{jacobian}
\end{equation}
The trace is now taken in functional (euclidean) space time. In terms
of the collective bilocals, the action for the non-linear sigma model
reads

\begin{equation}
S_{eff}=N\left[-\frac{1}{2}\mathrm{Tr}\ln\psi+\int d^{d}x\left(-\frac{1}{2}\lim_{y\rightarrow x}\partial^{2}\psi_{xy}+\frac{1}{2}\alpha_{x}\psi_{xx}-\frac{1}{2\lambda}\alpha_{x}\right)\right].
\end{equation}

\noindent The large $N$ saddle point equations of motion can be obtained
by varying the action above with respect to the bilocals and the Lagrange
multiplier.

With a large $N$ translational invariant ansatz

\[
\psi_{xy}=\int\frac{d^{d}k}{\left(2\pi\right)^{d}}e^{ik\left(x-y\right)}\psi_{k},
\]

\noindent we vary the effective action with respect to the bilocals
and find that

\begin{equation}
\psi_{k}^{0}=\frac{1}{k^{2}+\alpha},
\end{equation}
or in coordinate space

\begin{equation}
\psi_{xy}=\int\frac{d^{d}k}{\left(2\pi\right)^{d}}\frac{e^{ik\left(x-y\right)}}{k^{2}+\alpha}.
\end{equation}
Likewise, varying the effective action with respect to $\alpha_{x}$
leads us to

\begin{equation}
\psi_{xx}=\frac{1}{\lambda}.
\end{equation}
We then have the gap equation

\begin{equation}
\int\frac{d^{d}k}{\left(2\pi\right)^{d}}\frac{1}{k^{2}+\alpha}=\frac{1}{\lambda}.
\end{equation}

\noindent Thus, from (\ref{regularization}), one has  \cite{Lang:1990ni,Giombi:2009wh}
\begin{equation}
\frac{1}{\left(4\pi\right)^{d/2}}\Gamma\left(1-\frac{d}{2}\right)\alpha^{\frac{d-2}{2}}=\frac{1}{\lambda}.
\end{equation}

\noindent For $d=3$,

\begin{equation}
\sqrt{\alpha}=-\frac{4\pi}{\lambda}\label{mass nl sigma}
\end{equation}

\noindent and the critical point is reached as $\lambda\to\infty$,
corresponding to the large-$N$ conformal background configuration:

\begin{equation}
\psi_{k}^{0}=\frac{1}{k^{2}}.
\end{equation}

\noindent To generate $1/N$ corrections, we expand about this large-$N$
background\footnote{In \cite{Lang:1990ni}, Lang and Ruhl introduce an imaginary $\tilde{\alpha}$. Here, we will keep $\tilde{\alpha}$ real, to ensure that the coefficient of the two point function is positive; this is also the choice of Giombi and Yin \cite{Giombi:2009wh} and the later work of Leonhardt and Ruhl \cite{Leonhardt:2003du}.}:

\begin{flalign}
\alpha & =0+\frac{1}{\sqrt{N}}\tilde{\alpha}\left(x\right)\label{eq:alpha fluctuations}\\
\psi_{xy} & =\psi_{xy}^{0}+\frac{1}{\sqrt{N}}\eta_{xy}.\label{eq:bilocal fluctuations}
\end{flalign}

\noindent and insert (\ref{eq:alpha fluctuations}) and (\ref{eq:bilocal fluctuations})
into the effective action, keeping only terms quadratic in the fields.
This quadratic effective action reads:

\begin{equation}
S_{eff}^{\left(2\right)}=\frac{1}{4}\mathrm{Tr}\left(\psi_{0}^{-1}\tilde{\eta}\psi_{0}^{-1}\tilde{\eta}\right)-\frac{1}{4}\mathrm{Tr}\left(\tilde{\alpha}\psi_{0}\tilde{\alpha}\psi_{0}\right)
\end{equation}

\noindent after a shift of the bilocal fields defined by $\tilde{\eta}=\psi_{0}\tilde{\alpha}\psi_{0}+\eta$
which decouples the $\eta$ and $\tilde{\alpha}$ fields \cite{deMelloKoch:2011}.

We move into momentum space by writing

\begin{flalign}
\tilde{\eta}_{xy} & =\int\frac{d^{d}k_{1}}{\left(2\pi\right)^{d/2}}\int\frac{d^{d}k_{2}}{\left(2\pi\right)^{d/2}}e^{ik_{1}x}e^{ik_{2}y}\tilde{\eta}_{k_{1}k_{2}}\\
\tilde{\alpha}_{x} & =\int\frac{d^{d}k}{\left(2\pi\right)^{d/2}}e^{ikx}\tilde{\alpha}_{k}.
\end{flalign}
The quadratic effective action then becomes

\begin{flalign}
S_{eff}^{\left(2\right)} & =\frac{1}{4}\int d^{d}k_{1}\int d^{d}k_{2}\tilde{\eta}_{k_{1}k_{2}}\left(\psi_{0}^{-1}\right)_{k_{1}}\left(\psi_{0}^{-1}\right)_{k_{2}}\tilde{\eta}_{-k_{2},-k_{1}}\nonumber \\
 & -\frac{1}{4}\int d^{d}k_{1}\tilde{\alpha}_{k_{1}}\left(\int\frac{d^{d}p}{\left(2\pi\right)^{d}}\psi_{p}^{0}\psi_{k_{1}+p}^{0}\right)\tilde{\alpha}_{-k_{1}}.
\end{flalign}
Since \cite{Lang:1990ni,Das:2003vw}

\begin{flalign}
\int\frac{d^{d}p}{\left(2\pi\right)^{d}}\frac{1}{p^{2}}\frac{1}{\left(k-p\right)^{2}} & =-\frac{\left(k^{2}\right)^{\frac{d}{2}-2}\pi\Gamma\left(\frac{d}{2}-1\right)}{\left(4\pi\right)^{d/2}\sin\left(\frac{\pi d}{2}\right)\Gamma\left(d-2\right)}\nonumber \\
 & =\frac{1}{8\left|k\right|},\qquad d=3.\label{loop diagram}
\end{flalign}

\noindent
we can then write the quadratic effective action as

\begin{equation}
S_{eff}^{\left(2\right)}=\frac{1}{4}\int d^{d}k_{1}\int d^{d}k_{2}\tilde{\eta}_{k_{1}k_{2}}k_{1}^{2}k_{2}^{2}\tilde{\eta}_{-k_{2},-k_{1}}-\frac{1}{4}\int d^{d}k_{1}\tilde{\alpha}_{k_{1}}\left(\frac{1}{8\left|k\right|}\right)\tilde{\alpha}_{-k_{1}}.
\end{equation}
Therefore, the propagators \textendash{} which can be read off from
the quadratic effective action \textendash{} are

\begin{equation}
\left\langle \tilde{\eta}_{k_{1}k_{2}}\tilde{\eta}_{p_{1}p_{2}}\right\rangle =\frac{2}{k_{1}^{2}k_{2}^{2}}\delta\left(k_{2}+p_{2}\right)\delta\left(k_{1}+p_{1}\right)\label{eq:non linear sigma model eta eta propagator}
\end{equation}
and

\begin{equation}
\left\langle \tilde{\alpha}_{k_{1}}\tilde{\alpha}_{k_{2}}\right\rangle =-16\left|k_{1}\right|\delta\left(k_{1}+k_{2}\right).\label{eq: non-linear sigma model alpha alpha propagator}
\end{equation}
In coordinate space, the above propagators are

\begin{flalign}
\left\langle \eta_{x_{1}x_{2}}\eta_{x_{3}x_{4}}\right\rangle  & =\left(\frac{2^{d-2}}{\left(4\pi\right)^{d/2}}\Gamma\left(\frac{d}{2}-1\right)\right)^{2}\left(\left(x_{13}^{2}\right)^{1-\frac{d}{2}}\left(x_{24}^{2}\right)^{1-\frac{d}{2}}+\left(x_{14}^{2}\right)^{1-\frac{d}{2}}\left(x_{23}^{2}\right)^{1-\frac{d}{2}}\right)\nonumber \\
 & \rightarrow\left(\frac{1}{4\pi}\right)^{2}\left(\frac{1}{\left(x_{13}^{2}\right)^{1/2}}\frac{1}{\left(x_{24}^{2}\right)^{1/2}}+\frac{1}{\left(x_{14}^{2}\right)^{1/2}}\frac{1}{\left(x_{23}^{2}\right)^{1/2}}\right)\label{eq:spacetime correlation function for eta eta}
\end{flalign}
and

\begin{flalign}
\left\langle \tilde{\alpha}_{x_{1}}\tilde{\alpha}_{x_{2}}\right\rangle  & =2\left[\frac{\left(4\pi\right)^{d/2}\sin\left(\frac{\pi d}{2}\right)\Gamma\left(d-2\right)}{\pi\Gamma\left(\frac{d}{2}-1\right)}\right]\int\frac{d^{d}k}{\left(2\pi\right)^{d}}e^{ik\left(x_{1}-x_{2}\right)}\left(k^{2}\right)^{2-\frac{d}{2}}\nonumber \\
 & =2^{5}\left(\frac{\sin\left(\frac{\pi d}{2}\right)}{\pi}\right)\frac{\Gamma\left(d-2\right)}{\Gamma\left(\frac{d}{2}-1\right)\Gamma\left(\frac{d}{2}-2\right)}\frac{1}{\left(x_{12}^{2}\right)^{2}}\nonumber \\
 & \rightarrow \frac{16}{\pi^{2}}\frac{1}{\left(x_{12}^{2}\right)^{2}}\qquad d=3.
\end{flalign}

\noindent It follows that the conformal scaling dimension of the Lagrange
multiplier field is, as expected, $\Delta=2$. However, it is clear
that the scaling dimension of the local field $\eta_{xx}$ is given
by $\Delta=1$ - this can be seen by setting $x_{1}=x_{2}$ and $x_{3}=x_{4}$
in (\ref{eq:spacetime correlation function for eta eta}).

In conclusion, the non linear sigma model points to two types of excitations
at the IR critical point: a $\Delta=2$ state, and bilocal excitations
identical to the free theory which contain a local $\Delta=1$ state.
We will evidence an entirely similar structure in the $\frac{\lambda}{N}(\phi^{a}\phi^{a})^{2}$
theory propagator in the next section.

\section{$\left(\phi^{2}\right)^{2}$ in two-time bilocal approach }

We now consider the path integral formulation of the $\left(\phi^{2}\right)^{2}$
theory (\ref{lagrangian}), in terms of the two-time (covariant) collective
$O(N)$ invariant bilocals:\footnote{Recall the notation $x\equiv x^{\mu}=(t,\vec{x})$, and similarly
for momentum, $k\equiv k^{\mu}=(E,\vec{k})$. Our signature is $(+,-,-)$}

\begin{equation}
\psi_{xy}=\sum_{a=1}^{N}\phi^{a}\left(x\right)\phi^{a}\left(y\right).
\end{equation}

\noindent Including the Jacobian (\ref{jacobian}) resulting from
the change of variables to bilocal fields, the effective action for
the $O(N)$ $\left(\phi^{2}\right)^{2}$ vector theory in Minkowski
spacetime, is

\begin{equation}
S_{eff}=N\int d^{d}x\left[\frac{1}{2}\left(-\lim_{y\rightarrow x}\partial_{y}^{2}\psi_{xy}\right)-\frac{1}{2}m^{2}\psi_{xx}-\frac{\lambda}{4!}\left(\psi_{xx}\right)^{2}\right]-\frac{Ni}{2}\mathrm{Tr}\ln\psi
\end{equation}

\noindent The large-$N$ background is now a saddle point solution
with the translational invariant ansatz

\[
\psi_{xy}=\int\frac{d^{d}k}{\left(2\pi\right)^{d}}e^{ik\left(x-y\right)}\psi_{k},
\]

\noindent and it yields

\begin{equation}
\psi_{k}^{0}=\frac{i}{k^{2}-m^{2}-\frac{\lambda}{6}\int\frac{d^{d}k'}{\left(2\pi\right)^{d}}\psi_{k'}^{0}}.
\end{equation}

\noindent The solution to the ensuing gap equation has been described
in detail in Section (\ref{hamiltonian}), as well as the approach
and identification of the IR critical point. At the critical point,
the large $N$ background takes the conformal form:

\begin{equation}
\psi_{k}^{0}=\frac{i}{k^{2}}.
\end{equation}
We expand about this large-$N$ background and write

\begin{equation}
\psi_{xy}=\psi_{xy}^{0}+\frac{1}{\sqrt{N}}\eta_{xy}
\end{equation}
The quadratic effective action can be written as

\begin{equation}
S_{eff}^{\left(2\right)}=\frac{i}{4}\mathrm{Tr}\left(\psi_{0}^{-1}{\eta}\psi_{0}^{-1}{\eta}\right)-\frac{\lambda}{4!}\int{d^{d}x}\eta_{xx}^{2}
\end{equation}

\noindent or

\begin{equation}
iS_{eff}^{\left(2\right)}=-\frac{1}{2}\int d^{d}k_{1}\int d^{d}k_{2}\int d^{d}k_{3}\int d^{d}k_{4}\eta_{k_{1}k_{2}}\hat{O}_{k_{1}k_{2};k_{3}k_{4}}\eta_{k_{3}k_{4}},
\end{equation}
with

\begin{equation}
\hat{O}_{k_{1}k_{2};k_{3}k_{4}}=\frac{1}{2}\psi_{k_{3}}^{0^{-1}}\psi_{k_{4}}^{0^{-1}}\delta\left(k_{2}+k_{3}\right)\delta\left(k_{1}+k_{4}\right)+\frac{2i\lambda}{4!}\frac{1}{\left(2\pi\right)^{d}}\delta\left(k_{1}+k_{2}+k_{3}+k_{4}\right).\label{Defintion of O}
\end{equation}

\noindent The Fourier transformation has been defined as:

\begin{equation}
{\eta}_{xy}=\int\frac{d^{d}k_{1}}{\left(2\pi\right)^{d/2}}\int\frac{d^{d}k_{2}}{\left(2\pi\right)^{d/2}}e^{-ik_{1}x}e^{-ik_{2}y}{\eta}_{k_{1}k_{2}}.
\end{equation}

\noindent The inversion of this operator to yield the (collective
field bilocal) propagator has been described in \cite{deMelloKoch:1996mj}.
It corresponds to a Bethe-Salpeter equation for quartic correlators
of the underlying vector theory. The answer is:\footnote{We freely use the property that $\psi_{k}^0=\psi_{-k}^0$}

\begin{flalign}
\hat{O}_{k_{1}k_{2};p_{1}p_{2}}^{-1}=2\psi_{p_{1}}^{0}\psi_{p_{2}}^{0}\delta\left(k_{1}+p_{2}\right)\delta\left(k_{2}+p_{1}\right)\nonumber \\
+\frac{-\frac{i\lambda}{3}\frac{1}{\left(2\pi\right)^{d}}\psi_{k_{1}}^{0}\psi_{k_{2}}^{0}\psi_{p_{1}}^{0}\psi_{p_{2}}^{0}}{1+\frac{i\lambda}{6}\frac{1}{\left(2\pi\right)^{d}}\int d^{d}k_{1}\int d^{d}k_{2}\delta\left(k_{1}+k_{2}-p_{1}-p_{2}\right)\psi_{k_{1}}^{0}\psi_{k_{2}}^{0}}\delta\left(k_{1}+k_{2}+p_{1}+p_{2}\right).\label{eq:bilocal propagator in d dimensions}
\end{flalign}
The integral in the denominator follows from its euclidean version
(\ref{loop diagram}) with result (in $3d$):
\begin{equation}
\frac{1}{\left(2\pi\right)^{3}}\int d^{3}k_{1}\int d^{3}k_{2}\delta\left(k_{1}+k_{2}-p_{1}-p_{2}\right)\psi_{k_{1}}^{0}\psi_{k_{2}}^{0}=-\frac{i}{8}\frac{1}{\left|p_{1}+p_{2}\right|_{E}}.\label{eq:loop integral}
\end{equation}
As a result, the $3d$ two-time collective bilocal propagator is

\begin{flalign}
\hat{O}_{k_{1}k_{2};p_{1}p_{2}}^{-1}=2\psi_{p_{1}}^{0}\psi_{p_{2}}^{0}\delta\left(k_{1}+p_{2}\right)\delta\left(k_{2}+p_{1}\right)\nonumber \\
+\frac{-\frac{i\lambda}{3}\frac{1}{\left(2\pi\right)^{3}}\psi_{k_{1}}^{0}\psi_{k_{2}}^{0}\psi_{p_{1}}^{0}\psi_{p_{2}}^{0}}{1+\frac{\lambda}{48}\frac{1}{\left|p_{1}+p_{2}\right|_{E}}}\delta\left(k_{1}+k_{2}+p_{1}+p_{2}\right).\label{eq:bilocal propagator}
\end{flalign}

\noindent
The bilocal propagator consists of a free, disconnected piece (identical
to the UV critical point) associated with the free propagation of
two underlying scalars, and of a s-channel scattering of a composite
state. The mass shell condition is as usual obtained by identifying
the pole of the propagator, after removal of external legs. The pole
condition in (\ref{eq:bilocal propagator}), is

\begin{equation}
1=-\frac{\lambda}{48}\frac{1}{\left|p_{1}+p_{2}\right|_{E}} \label{sign},
\end{equation}
or

\begin{equation}
E^{2}-\left(\vec{p}_{1}+\vec{p}_{2}\right)^{2}=(E_1+E_2)^{2}-\left(\vec{p}_{1}+\vec{p}_{2}\right)^{2}=-\frac{\lambda^{2}}{48^{2}}.\label{b state energy}
\end{equation}

\noindent
It may be of concern that (\ref{sign}) was obtained with background large $N$ exact massless propagators whereas for the pole condition $\lambda$ was kept large, but finite. This is unfounded, as the result (\ref{eq:loop integral}) is finite and does not require regularization.  As a matter of fact, it has been obtained in closed form with massive propagators in the first of \cite{Lang:1990ni}, from which one can obtain:

\begin{equation}
\left|p_{1}+p_{2}\right|_{E}= -\frac{\lambda}{48} - \frac{4}{\pi} \frac{m^2}{\lambda} + ...
\end{equation}

\noindent
The limit $\lambda\to\infty$ results in the finite (independent of $\lambda$) propagator presented below.  

We also remark that equation (\ref{sign}) requires $\lambda<0$, i.e., an attractive quartic potential. This is the case if to agree with the non linear sigma model approach, as it can be seen by comparing  (\ref{mass}) with (\ref{mass nl sigma}). It is also "natural" in the standard auxiliary field formulation of the quartic theory. A close examination of, for instance, \cite{ZinnJustin}, shows that for the associated gaussian integral to be well defined, either $\lambda<0$ or the auxiliary field is imaginary. This is related to the discussion in the footnote just before equation (\ref{eq:alpha fluctuations}). The discussion of the sign of the coupling  is reminiscent of the Gross-Neveu model \cite{Gross:1974jv}, although in this case it applies to fermionic theories.

Returning to (\ref{eq:bilocal propagator}), we note that
at the infra-red fixed point ($\lambda\to\infty$) the propagator
takes the finite form:

\begin{flalign}
\hat{O}_{k_{1}k_{2};p_{1}p_{2}}^{-1}=2\frac{i}{p_{1}^{2}}\frac{i}{p_{2}^{2}}\delta\left(k_{1}+p_{2}\right)\delta\left(k_{2}+p_{1}\right)\nonumber \\
-\frac{i}{k_{1}^{2}}\frac{i}{k_{2}^{2}}\frac{16i\left|p_{1}+p_{2}\right|_{E}}{\left(2\pi\right)^{3}}\frac{i}{p_{1}^{2}}\frac{i}{p_{2}^{2}}\delta\left(k_{1}+k_{2}+p_{1}+p_{2}\right).\label{finite propagator}
\end{flalign}

This result for the bilocal propagator is in direct agreement with
the non-linear sigma model results (\ref{eq:non linear sigma model eta eta propagator})
and (\ref{eq: non-linear sigma model alpha alpha propagator}), up
to leg-factors in the $\Delta=2$ channel, confirming the identification
of the intermediate state as a $\Delta=2$ state at criticality. Conformal invariance
and dimensional analysis dictates for such a state an infinite "pole" in the two point function. Equation (\ref{b state energy}) makes
precise how this limit is approached.\footnote{In conformal field theories, mass states are not in their Cartan subalgebras, but it is legitimate to discuss the approach to conformal criticality.} 

\section{From covariant bilocals to equal time bilocals}

It is of interest to consider the equations of motion satisfied by
the covariant bilocals fluctuations $\hat{O}\eta=0$. With $\hat{O}$
defined in \eqref{Defintion of O}, one has:

\begin{equation}
\psi_{k_{1}}^{{0}^{-1}}\psi_{k_{2}}^{{0}^{-1}}\eta_{k_{1}k_{2}}=-\frac{4i\lambda}{4!}\frac{1}{\left(2\pi\right)^{d}}\int d^{d}k\eta_{k,k_{1}+k_{2}-k}.\label{eq:integral equation for covariant bilocals}
\end{equation}
or 
\begin{equation}
{k_{1}}^{2}{k_{2}}^{2}\eta_{k_{1}k_{2}}=\frac{i\lambda}{6}\frac{1}{\left(2\pi\right)^{d}}\int d^{d}k\eta_{k,k_{1}+k_{2}-k}.\label{integral equation}
\end{equation}

\noindent In coordinate space,

\begin{equation}
\partial_{x}^{2}\partial_{y}^{2}\eta_{xy}=\frac{i\lambda}{6}\delta(x-y)\eta_{xx}.
\end{equation}

\noindent Rewriting (\ref{integral equation}) as

\begin{equation}
\eta_{k_{1}k_{2}}=\frac{i\lambda}{6}\frac{1}{\left(2\pi\right)^{d}}\frac{1}{k_{1}^{2}}\frac{1}{k_{2}^{2}}\int d^{d}k\eta_{k,k_{1}+k_{2}-k}.\label{eq:int eq}
\end{equation}

\noindent One can show that

\begin{equation}
\eta_{k_{1}k_{2}}=\frac{\alpha_{k_{1}+k_{2}}}{k_{1}^{2}k_{2}^{2}}.\label{mom eigfunctions}
\end{equation}

\noindent is a solution of (\ref{eq:int eq}) for arbitrary $\alpha_k$,
provided

\begin{equation}
1=-\frac{4i\lambda}{4!}\frac{1}{\left(2\pi\right)^{d}}\int d^{d}k_{1}\psi_{k_{1}}^{0}\psi_{k_{1}-p_{1}-p_{2}}^{0}\label{eq:two pole condition}
\end{equation}

\noindent is satisfied. We recognize this equation as the pole condition
of the previous section. The $\eta_{k_{1}k_{2}}$ of the form (\ref{mom eigfunctions}) are nothing but the (momentum space) bound state eigenfunctions.  



We have so far two descriptions \textit{viz.} one in terms of the
covariant bilocals and the other one in terms of the single time bilocals.
The results in the two descriptions look superficially different.
However, the single time equations (\ref{eq:single time equation for the fluctuations})
and (\ref{ham pole condition}) should correspond to equations (\ref{eq:int eq})
and (\ref{eq:two pole condition}) respectively. In the following
we show that they are indeed equivalent.

Recall that 

\begin{flalign}
{\eta}_{xy} & ={\eta}_{t_{x},\vec{x};t_{y},\vec{y}}=\int\frac{d^{d}k_{1}}{\left(2\pi\right)^{d/2}}\int\frac{d^{d}k_{2}}{\left(2\pi\right)^{d/2}}e^{-ik_{1}x}e^{-ik_{2}y}{\eta}_{k_{1}k_{2}}\nonumber \\
 & =\int\frac{dE_{1}d^{d-1}\vec{k_{1}}}{\left(2\pi\right)^{d/2}}\int\frac{dE_{2}d^{d-1}\vec{k_{2}}}{\left(2\pi\right)^{d/2}}e^{-iE_{1}t_{x}+i\vec{k_{1}}\vec{x}}e^{-iE_{2}t_{y}+i\vec{k_{2}}\vec{y}}{\eta}_{E_{1}\vec{k_{1}};E_{2}\vec{k_{2}}}.
\end{flalign}



\noindent
Equal time bilocals are obtained from covariant bilocals by setting
$t_{x}=t_{y}=t$ or, equivalently, they only depend on $E=E_{1}+E_{2}$
and as such can be obtained by integration of an intermediate energy
variable (in this case, $E_{1}-E_{2}$).

In the appendix we establish the result:

\begin{multline}
\int\frac{dE}{2\pi}\frac{1}{E^{2}-\vec{k}^{2}+i\epsilon}\frac{1}{\left(E-E_{p}\right)^{2}-\left(\vec{k}-\vec{p}\right)^{2}+i\epsilon}=-\frac{i}{2}\\
\times\frac{1}{E_{p}^{2}-\left(\left|\vec{k}\right|+\left|\vec{k}-\vec{p}\right|\right)^{2}}\left(\frac{1}{\left|\vec{k_{}}\right|}+\frac{1}{\left|\vec{k}-\vec{p}\right|}\right).\label{eq:integral over E}
\end{multline}

\noindent
In other words,


\begin{flalign}
&\int\frac{d^{d-1}\vec{k}}{\left(2\pi\right)^{d-1}}\frac{1}{E_{p}^{2}-\left(\left|\vec{k}\right|+\left|\vec{p}-\vec{k}\right|\right)^{2}}\left(\frac{1}{\left|\vec{k}\right|}+\frac{1}{\left|\vec{p}-\vec{k}\right|}\right) \nonumber\\
= 2i &\int \frac{d^{d}k_{1}}{{\left(2\pi\right)^{d}}}\frac{1}{k^{2}}\frac{1}{\left(k-p\right)^{2}} \label{two to one integral}
\end{flalign}

\noindent
Note that $p=p^{\mu}=(E_p,\vec{p})$, with $E_p$ otherwise arbitrary. These results establish the equivalence of (\ref{eq:single time equation for the fluctuations}),
(\ref{ham pole condition}) and equations (\ref{eq:int eq}), (\ref{eq:two pole condition})
respectively. Explicitly,  it follows immediately that the pole condition (\ref{eq:two pole condition})

\begin{equation}
1=\frac{i\lambda}{6}\frac{1}{\left(2\pi\right)^{d}}\int d^{d}k_{1}\frac{1}{k^{2}}\frac{1}{\left(k-p_{1}-p_{2}\right)^{2}}\nonumber 
\end{equation}

\noindent
is equivalent to 

\begin{equation}
1=\frac{\lambda}{12}\int\frac{d^{d-1}\vec{k}}{\left(2\pi\right)^{d-1}}\frac{1}{E_{p_{1}+p_{2}}^{2}-\left(\left|\vec{k}\right|+\left|\vec{p_{1}}+\vec{p_{2}}-\vec{k}\right|\right)^{2}}\left(\frac{1}{\left|\vec{k}\right|}+\frac{1}{\left|\vec{p_{1}}+\vec{p_{2}}-\vec{k}\right|}\right),
\end{equation}
which is nothing but the Hamiltonian pole condition (\ref{ham pole condition}).
The solution is the bound state with dispersion (\ref{bound state energy}).
At the critical point $\lambda\to\infty$ this state has been identified
in the path integral as a $\Delta=2$ conformal field, both directly
in the non-linear sigma model treatment of the IR critical point,
and for the $\frac{\lambda}{N}(\phi^{a}\phi^{a})^{2}$ theory.

\section{Potential scattering, the fate of the $\Delta=1$ state and the $\Delta=2$ bound state}


There is a puzzle when one considers the results of the previous sections.
Both bilocal propagators (\ref{eq:non linear sigma model eta eta propagator}-\ref{eq: non-linear sigma model alpha alpha propagator}) and (\ref{eq:bilocal propagator}) display disconnected diagrams identical
to those of the (free) UV critical point, in addition to the s-channel
$\Delta=2$ bound state. One would expect the states associated with
these disconnected diagrams to include a $\Delta=1$ boundary state,
which should not be present at the IR critical point. On the other
hand, it is not clear how these free states are solutions of the Hamiltonian
equations of motion (\ref{scattering equation}), certainly when written in the form (\ref{eq:single time equation for the fluctuations}). But
these are the states needed to \char`\"{}build\char`\"{} the bulk.

The answer is that the most general scattering state solution to the equation


\begin{equation}
E^2 \eta_{\vec{k}_{1}\vec{k}_{2}}=\frac{1}{4}\left(\psi_{\vec{k}_{1}}^{0^{-1}}+\psi_{\vec{k}_{2}}^{0^{-1}}\right)^{2}\eta_{\vec{k}_{1}\vec{k}_{2}}+\frac{\lambda}{6}\left(\psi_{\vec{k}_{1}}^{0}+\psi_{\vec{k}_{2}}^{0}\right)\int\frac{d^{d-1}\vec{l}}{\left(2\pi\right)^{d-1}}\eta_{\vec{k}_{1}+\vec{k}_{2}-\vec{l},\vec{l}},,
\end{equation}

\noindent
best regarded as a (relativistic) potential scattering problem, is given by\footnote{Here, $\vec{p}=\vec{k}_{1}+\vec{k}_{2}$.}


\begin{equation}
\eta_{\vec{k_1},\vec{p}-\vec{k_1}}=\varphi_{\vec{k_1},\vec{p}-\vec{k_1}}+\frac{   \frac{\lambda}{12} \left(\frac{1}{\left|\vec{k_1}\right|}+\frac{1}{\left|\vec{p}-\vec{k_1}\right|}\right)}{E^{2}-\left(\left|\vec{k_1}\right|+\left|\vec{p}-\vec{k_1}\right|\right)^{2}}\int\frac{d^{2}\vec{l}}{\left(2\pi\right)^{2}}\eta_{\vec{l}.\vec{p}-\vec{l}},\label{eq:complete solution}
\end{equation}
where $\varphi_{\vec{k}_{1}\vec{k}_{2}}$ solves the free equation
of motion, i.e.,

\begin{equation}
E^{2}=\left(\left|\vec{p}_{1}\right|+\left|\vec{p}_{2}\right|\right)^{2},\qquad\varphi_{\vec{k}_{1}\vec{k}_{2}}\sim\delta\left(\vec{k}_{1}-\vec{p}_{1}\right)\delta\left(\vec{k}_{2}-\vec{p}_{2}\right)
\end{equation}


\noindent
Integrating both sides of (\ref{eq:complete solution}) leads to

\begin{equation}
\int \frac{d^{2}\vec{k}}{(2\pi)^2}\eta_{\vec{k},\vec{p}-\vec{k}}=\int \frac{d^{2}\vec{k}}{(2\pi)^2}\varphi_{\vec{k},\vec{p}-\vec{k}}- \frac{\lambda}{48\left| p \right|_E}\int \frac{d^{2}\vec{k}}{(2\pi)^2} \eta_{\vec{k},\vec{p}-\vec{k}},
\end{equation}
where in order to arrive at the final expression we have made use
of (\ref{eq:loop integral}) and (\ref{two to one integral}). Note that

\begin{equation}
p^{\mu} = (|\vec{p_1}|+|\vec{p_2}|, \vec{p_1}+\vec{p_2}),
\end{equation}



\noindent
with $p_E$ the corresponding euclidean $3$-vector. Thus,

\begin{equation}
\int \frac{d^{2}\vec{k}}{(2\pi)^2} \eta_{\vec{k},\vec{p}-\vec{k}}=\frac{1}{1+\frac{\lambda}{48\left| p \right|_E}}\int \frac{d^{2}\vec{k}}{(2\pi)^2}\varphi_{\vec{k},\vec{p}-\vec{k}}.
\end{equation}

\noindent
This can be substituted back into (\ref{eq:complete solution}), resulting in the equivalent expression

\begin{equation}
\int \frac{d^{2}\vec{k}}{(2\pi)^2} \eta_{\vec{k},\vec{p}-\vec{k}}=\left( 1 - \frac{\lambda}{48\left| p \right|_E} \frac{1}{1+\frac{\lambda}{48\left| p \right|_E}}   \right)  \int \frac{d^{2}\vec{k}}{(2\pi)^2}\varphi_{\vec{k},\vec{p}-\vec{k}}.
\end{equation}

\noindent
Either way, at the infra-red conformal point $\lambda\rightarrow\infty$, 

\begin{equation}
\eta_{xx}\sim\int \frac{d^{2}\vec{k}}{(2\pi)^2}\eta_{\vec{k},\vec{p}-\vec{k}} \to 0, \label{psizero}
\end{equation}

\noindent
and this state with $\Delta=1$ is removed from the spectrum.

We provide a further check by examining the propagator $\left\langle \eta_{xx}\eta_{yy}\right\rangle$, in the path integral approach: 

\begin{multline}
\left\langle \eta_{xx}\eta_{yy}\right\rangle =\int\frac{d^{3}k_{1}}{\left(2\pi\right)^{3/2}}\frac{d^{3}k_{2}}{\left(2\pi\right)^{3/2}}\frac{d^{3}p_{1}}{\left(2\pi\right)^{3/2}}\frac{d^{3}p_{2}}{\left(2\pi\right)^{3/2}}
e^{-ix\left(k_{1}+k_{2}\right)}e^{-iy\left(p_{1}+p_{2}\right)}\hat{O}_{k_{1}k_{2};p_{1}p_{2}}^{-1}\\
=\int\frac{d^{3}k_{1}}{\left(2\pi\right)^{3/2}}\frac{d^{3}k_{2}}{\left(2\pi\right)^{3/2}}\frac{d^{3}p_{1}}{\left(2\pi\right)^{3/2}}\frac{d^{3}p_{2}}{\left(2\pi\right)^{3/2}}e^{-ix\left(k_{1}+k_{2}\right)}e^{-iy\left(p_{1}+p_{2}\right)}  \times \\
\left(-2\frac{1}{p_{1}^{2}}\frac{1}{p_{2}^{2}}\delta\left(k_{1}+p_{2}\right)\delta\left(k_{2}+p_{1}\right)
-\frac{16i\left|p_{1}+p_{2}\right|_{E}}{\left(2\pi\right)^{3}}
\frac{1}{k_{1}^{2}}\frac{1}{k_{2}^{2}}\frac{1}{p_{1}^{2}}\frac{1}{p_{2}^{2}}\delta\left(k_{1}+k_{2}+p_{1}+p_{2}\right)\right)\\
=-2\int\frac{d^{3}k_{1}}{\left(2\pi\right)^{3}}\frac{d^{3}k_{2}}{\left(2\pi\right)^{3}}\frac{e^{ik_{1}\left(y-x\right)}}{k_{1}^{2}}\frac{e^{ik_{2}\left(y-x\right)}}{k_{2}^{2}}\\
-16i\int\frac{d^{3}k_{1}}{\left(2\pi\right)^{3}}\frac{d^{3}k_{2}}{\left(2\pi\right)^{3}}\left(\int \frac{d^{3}p_{1}}{\left(2\pi\right)^{3}}\frac{1}{p_{1}^{2}}\frac{1}{\left(p_{1}+k_{1}+k_{2}\right)^{2}}\right)\frac{e^{ik_{1}\left(y-x\right)}}{k_{1}^{2}}\frac{e^{ik_{2}\left(y-x\right)}}{k_{2}^{2}}\left|k_{1}+k_{2}\right|_{E}.
\end{multline}

From \eqref{eq:loop integral}, we have

\begin{equation}
\int\frac{d^{3}p_{1}}{\left(2\pi\right)^{3}}\frac{1}{p_{1}^{2}}\frac{1}{\left(p_{1}+k_{1}+k_{2}\right)^{2}}=\frac{i}{8\left|k_{1}+k_{2}\right|_{E}}.
\end{equation}

Accordingly,

\begin{multline}
-16i\int\frac{d^{3}k_{1}}{\left(2\pi\right)^{3}}\frac{d^{3}k_{2}}{\left(2\pi\right)^{3}}\left(\int \frac{d^{3}p_{1}}{\left(2\pi\right)^{3}}\frac{1}{p_{1}^{2}}\frac{1}{\left(p_{1}+k_{1}+k_{2}\right)^{2}}\right)\frac{e^{ik_{1}\left(y-x\right)}}{k_{1}^{2}}\frac{e^{ik_{2}\left(y-x\right)}}{k_{2}^{2}}\left|k_{1}+k_{2}\right|_{E}=\\
-16i\int\frac{d^{3}k_{1}}{\left(2\pi\right)^{3}}\frac{d^{3}k_{2}}{\left(2\pi\right)^{3}}\left(\frac{i}{8\left|k_{1}+k_{2}\right|_{E}}\right)\frac{e^{ik_{1}\left(y-x\right)}}{k_{1}^{2}}\frac{e^{ik_{2}\left(y-x\right)}}{k_{2}^{2}}\left|k_{1}+k_{2}\right|_{E}\\
=-2\left(-1\right)\int\frac{d^{3}k_{1}}{\left(2\pi\right)^{3}}\frac{d^{3}k_{2}}{\left(2\pi\right)^{3}}\frac{e^{ik_{1}\left(y-x\right)}}{k_{1}^{2}}\frac{e^{ik_{2}\left(y-x\right)}}{k_{2}^{2}}.
\end{multline}

Thus,

\begin{flalign}
\left\langle \eta_{xx}\eta_{yy}\right\rangle  & =-2\int\frac{d^{3}k_{1}}{\left(2\pi\right)^{3}}\frac{d^{3}k_{2}}{\left(2\pi\right)^{3}}\frac{e^{ik_{1}\left(y-x\right)}}{k_{1}^{2}}\frac{e^{ik_{2}\left(y-x\right)}}{k_{2}^{2}}-2\left(-1\right)\int\frac{d^{3}k_{1}}{\left(2\pi\right)^{3}}\frac{d^{3}k_{2}}{\left(2\pi\right)^{3}}\frac{e^{ik_{1}\left(y-x\right)}}{k_{1}^{2}}\frac{e^{ik_{2}\left(y-x\right)}}{k_{2}^{2}}\nonumber \\
 & =0.
\end{flalign}

\noindent
How is one able to extract $\Delta=2$ correlators from the bilocal fields? This is suggested from the discussion after equation (\ref{finite propagator}) and the eigenfunctions (\ref{mom eigfunctions}):

\begin{equation}
\eta_{xy}=(\psi^0 \alpha \psi^0)_{xy} ,
\end{equation}

\noindent
where $\alpha$ is the $\Delta=2$ field. Inverting,

\begin{equation}
\alpha(x) \delta(x-y)= ({\psi^0}^{-1} \eta {\psi^0}^{-1})_{xy} \label{invert}
\end{equation}

\noindent
Note that there is no a priori guarantee from this definition that correlators calculated with the right hand side of the above equation will always appear multiplied by a delta function, allowing one to extract $\alpha$ correlators. We will show this to be the case. 

For convenience, we change to an euclidean signature (with the Jacobian remaining unchanged, this simply requires $\lambda \to -i \lambda$ and $\psi^0_k=1/k^2$) so that the critical bilocal propagator (\ref{finite propagator}) takes the form:

\begin{flalign}
\hat{O^E}_{k_{1}k_{2};p_{1}p_{2}}^{-1}=2\frac{1}{p_{1}^{2}}\frac{1}{p_{2}^{2}}\delta\left(k_{1}+p_{2}\right)\delta\left(k_{2}+p_{1}\right)\nonumber \\
-\frac{1}{k_{1}^{2}}\frac{1}{k_{2}^{2}}\frac{16\left|p_{1}+p_{2}\right|_{E}}{\left(2\pi\right)^{3}}\frac{1}{p_{1}^{2}}\frac{1}{p_{2}^{2}}\delta\left(k_{1}+k_{2}+p_{1}+p_{2}\right).\label{euc finite propagator}
\end{flalign}

One has
 
 \begin{multline}
\left\langle ({\psi^0}^{-1} \eta {\psi^0}^{-1})_{x_1y_1} ({\psi^0}^{-1} \eta {\psi^0}^{-1})_{x_2y_2} \right\rangle \\
=\int\frac{d^{3}k_{1}}{\left(2\pi\right)^{3/2}}\frac{d^{3}k_{2}}{\left(2\pi\right)^{3/2}}\frac{d^{3}p_{1}}{\left(2\pi\right)^{3/2}}\frac{d^{3}p_{2}}{\left(2\pi\right)^{3/2}}e^{i k_{1} x_1} e^{ik_{2}y_1} e^{i p_{1} x_2} e^{ip_{2}y_2}  
k_1^2k_2^2p_1^2p_2^2 \left\langle  \eta_{k_1 k_2} \eta_{p_1 p_2}        \right\rangle  
\end{multline}
 
\noindent
The contribution from the connected piece of the bilocal propagator is
 
 \begin{multline}
 \left\langle ({\psi^0}^{-1} \eta {\psi^0}^{-1})_{x_1y_1} ({\psi^0}^{-1} \eta {\psi^0}^{-1})_{x_2y_2} \right\rangle \\ = \delta(x_1-y_1)  \delta(x_2-y_2)  \int\frac{d^{3}p}{\left(2\pi\right)^{3}} e^{i p (y_{2}- y_1)} (- 16 |p| ) =
 \delta(x_1-y_1) \delta(x_2-y_2)  \left\langle \alpha(x_1) \alpha(x_2) \right\rangle
 \end{multline}

\noindent
in agreement with (\ref{eq: non-linear sigma model alpha alpha propagator}). For the contribution from the disconnected piece of the bilocal propagator, recall that we consistently use: 

\begin{equation}
\int d^dp (p^2)^\alpha e^{ipx} = \pi^{d/2} 2^{2\alpha+d} \frac{\Gamma(\alpha+d/2)}{\Gamma(-\alpha)}
 \left( x^2 \right)^{-d/2-\alpha}.
 \end{equation}

\noindent
With this definition, it is straightforward to check that the disconnected contributions are proportional to $1/\Gamma(-1)$ and hence vanish.\footnote{There is also a concept of orthogonality. One can easily show that 
$\left\langle \eta_{xx} ({\psi^0}^{-1} \eta {\psi^0}^{-1})_{x_1y_1} \right\rangle \sim 0 \quad \delta (x_1-y_1)   $, as it should be the case for two fields with different conformal dimensions. This follows from a cancellation, again, between the
contributions of the connected and disconnected pieces of the critical bilocal propagator  }   

\section{Summary and Outlook}

The $O(N)$ invariant $\lambda\phi^4$ theory in $3$ dimensions has been studied systematically in a $1/N$ expansion at its infrared critical point, both in the Hamiltonian as well as in the path integral formalism. This systematic $1/N$ expansion is generated through $O(N)$ invariant bilocals following the collective field theory method of Jevicki and Sakita \cite{Jevicki:1979mb}. The presence of bilocal scattering states in the spectrum of the theory with free dispersion relations needed to generate the bulk according to the map of \cite{Koch:2010cy} was established, and the nature of the $\Delta =1$ and $\Delta=2$ fields has been elucidated. These fields have different origins. The $\Delta=1$ field is part of the bilocal scattering states, and it has been explicitly demonstrated in this article that it is absent from the spectrum at criticality. Marginally away from criticality and in the large $N$ conformal background, the $\Delta=2$ state is identified with a negative energy squared s-channel bound state with a finite (independent of $\lambda$) two point function at criticality. 

A related simpler simpler model that can be used to provide physical intuition to the features identified in this article is the non-relativistic one dimensional quantum mechanics with an attractive delta function potential $V=v_0 \delta(x)$,$v_0 < 0$. As is well known, this system has scattering states with $E>0$ and one bound state. In the limit that $v_0 \to -\infty$, an argument entirely similar to the one leading to (\ref{psizero}) shows that $\psi(0)=0$. In other words, the particles are prevented from "falling into the (infinitely deep) well". Despite the bound state having infinite negative energy in this limit, the second quantized two point function is finite and independent of $v_0$, in analogy with the critical
bilocal propagator presented above.

The general picture that emerges is then clear, following from the general properties of the map \cite{Koch:2010cy}, and particularly from (\ref{z coord}). Of the states in the bulk, the $\Delta=1$ state $\eta_{\vec{x_1}\vec{x_1}}$ (it follows from 
(\ref{z coord}) that $\vec{x}_2\to\vec{x}_1$ corresponds to the boundary) is not present at the boundary. In terms of the bilocal description, these scattering states are prevented from reaching the boundary. At the boundary, as it
follows from the identification (\ref{invert}), a decoupled $\Delta=2$ state is present which originates from a bound state in the three dimensional field theory.     

Of immediate future interest is to include this state in the map of \cite{Koch:2010cy}. Furthermore, as a spin $2$, $\Delta=2$ state with exponential real time dependence, it deserves further study.
 
\section{Appendix}

In this appendix, we derive the result in (\ref{eq:integral over E}),
\textit{viz. }

\begin{flalign}
\int\frac{dE}{2\pi}\frac{1}{E^{2}-\vec{k}^{2}+i\epsilon}\frac{1}{\left(E-E_{p}\right)^{2}-\left(\vec{k}-\vec{p}\right)^{2}+i\epsilon}=-\frac{i}{2}\nonumber \\
\times\frac{1}{E_{p}^{2}-\left(\left|\vec{k}_{1}\right|+\left|\vec{k}_{2}\right|\right)^{2}}\left(\frac{1}{\left|\vec{k_{_{1}}}\right|}+\frac{1}{\left|\vec{k_{_{2}}}\right|}\right).
\end{flalign}
Define

\begin{equation}
f\left(E\right)=\frac{1}{E^{2}-\vec{k}^{2}+i\epsilon}\frac{1}{\left(E-E_{p}\right)^{2}-\left(\vec{k}-\vec{p}\right)^{2}+i\epsilon}.
\end{equation}

The poles of the integrand are at $E=\pm\left(\left|\vec{k}\right|-i\epsilon\right)$
and $E=E_{p}\pm\left(\left|\vec{k}-\vec{p}\right|-i\epsilon\right)$.
We will choose to close the contour along the LHP. As a result, we
need to compute the residues at $E=\left|\vec{k}\right|-i\epsilon$
and $E=E_{p}+\left|\vec{k}-\vec{p}\right|-i\epsilon$.

The residue at $E=\left|\vec{k}\right|-i\epsilon$ is

\begin{equation}
\mathrm{Res}\left[f\left(\left|\vec{k}\right|\right)\right]=\frac{1}{2\left|\vec{k}\right|}\frac{1}{\left|\vec{k}\right|-E_{p}-\left|\vec{k}-\vec{p}\right|}\frac{1}{\left|\vec{k}\right|-E_{p}+\vec{k}-\vec{p}}
\end{equation}
and the one at $E=E_{p}+\left|\vec{k}-\vec{p}\right|-i\epsilon$ yields

\begin{equation}
\mathrm{Res}\left[f\left(E_{p}+\left|\vec{k}-\vec{p}\right|\right)\right]=\frac{1}{2\left|\vec{k}-\vec{p}\right|}\frac{1}{E_{p}+\left|\vec{k}-\vec{p}\right|-\left|\vec{k}\right|}\frac{1}{E_{p}+\left|\vec{k}-\vec{p}\right|+\left|\vec{k}\right|}.
\end{equation}
Using the residue theorem, we have

\begin{flalign}
\int\frac{dE}{\left(2\pi\right)}f\left(E\right) & =-i\Biggl[\frac{1}{2\left|\vec{k}\right|}\frac{1}{\left|\vec{k}\right|-E_{p}-\left|\vec{k}-\vec{p}\right|}\frac{1}{\left|\vec{k}\right|-E_{p}+\left|\vec{k}-\vec{p}\right|}\nonumber \\
 & +\frac{1}{2\left|\vec{k}-\vec{p}\right|}\frac{1}{E_{p}+\left|\vec{k}-\vec{p}\right|-\left|\vec{k}\right|}\frac{1}{E_{p}+\left|\vec{k}-\vec{p}\right|+\left|\vec{k}\right|}\Biggl].\label{eq:int f(E) =00003D00003D00003D00003D}
\end{flalign}

We define $\vec{k}=\vec{k_{1}}$ and $\vec{p}-\vec{k}=\vec{k_{2}}$.
and symmetrize the RHS of (\ref{eq:int f(E) =00003D00003D00003D00003D}).
This leads us to the result

\begin{flalign}
\int\frac{dE}{\left(2\pi\right)}f\left(E\right)=-\frac{i}{4}\Biggl[\frac{1}{\left|\vec{k}_{1}\right|}\frac{1}{\left|\vec{k_{1}}\right|-E_{p}-\left|\vec{k}_{2}\right|}\frac{1}{\left|\vec{k}_{1}\right|-E_{p}+\left|\vec{k}_{2}\right|}\nonumber \\
+\frac{1}{\left|\vec{k}_{2}\right|}\frac{1}{\left|\vec{k_{2}}\right|+E_{p}-\left|\vec{k}_{1}\right|}\frac{1}{\left|\vec{k_{2}}\right|+E_{p}+\left|\vec{k}_{1}\right|}\nonumber \\
+\frac{1}{\left|\vec{k}_{2}\right|}\frac{1}{\left|\vec{k_{2}}\right|-E_{p}-\left|\vec{k}_{1}\right|}\frac{1}{\left|\vec{k}_{2}\right|-E_{p}+\left|\vec{k}_{1}\right|}+\frac{1}{\left|\vec{k}_{1}\right|}\frac{1}{\left|\vec{k_{1}}\right|+E_{p}-\left|\vec{k}_{2}\right|}\frac{1}{\left|\vec{k}_{1}\right|+E_{p}+\left|\vec{k}_{2}\right|}\Biggr].
\end{flalign}
After some trivial but tedious manipulations, we obtain

\begin{multline}
\int\frac{dE}{\left(2\pi\right)}f\left(E\right)=-\frac{i}{4}\frac{1}{\left(E_{p}^{2}-\left(\left|\vec{k}_{1}\right|-\left|\vec{k}_{2}\right|\right)^{2}\right)\left(E_{p}^{2}-\left(\left|\vec{k}_{1}\right|+\left|\vec{k}_{2}\right|\right)^{2}\right)}\\
\times\Biggl[\frac{2}{\left|\vec{k}_{1}\right|}\Biggl[\left(E_{p}^{2}+\left|\vec{k}_{1}\right|^{2}-\left|\vec{k}_{2}\right|^{2}\right)+\frac{2}{\left|\vec{k}_{2}\right|}\left(E_{p}^{2}+\left|\vec{k}_{2}\right|^{2}-\left|\vec{k}_{1}\right|^{2}\right)\Biggl]\\
=-\frac{i}{2}\frac{1}{\left(E_{p}^{2}-\left(\left|\vec{k}_{1}\right|-\left|\vec{k}_{2}\right|\right)^{2}\right)\left(E_{p}^{2}+\left(\left|\vec{k}_{1}\right|+\left|\vec{k}_{2}\right|\right)^{2}\right)}\left[E_{p}^{2}\left(\frac{1}{\left|\vec{k_{_{1}}}\right|}+\frac{1}{\left|\vec{k_{2}}\right|}\right)\right.\\
\left.-\left(\left|\vec{k}_{2}\right|^{2}-\left|\vec{k}_{1}\right|^{2}\right)\left(\frac{1}{\left|\vec{k_{_{1}}}\right|}-\frac{1}{\left|\vec{k_{_{2}}}\right|}\right)\right]=-\frac{i}{2}\frac{1}{\left(E_{p}^{2}-\left(\left|\vec{k}_{1}\right|-\left|\vec{k}_{2}\right|\right)^{2}\right)\left(E_{p}^{2}-\left(\left|\vec{k}_{1}\right|+\left|\vec{k}_{2}\right|\right)^{2}\right)}\\
=-\frac{i}{2}\frac{1}{\left(E_{p}^{2}-\left(\left|\vec{k}_{1}\right|-\left|\vec{k}_{2}\right|\right)^{2}\right)\left(E_{p}^{2}-\left(\left|\vec{k}_{1}\right|+\left|\vec{k}_{2}\right|\right)^{2}\right)}\left[\frac{E_{p}^{2}}{\left|\vec{k_{_{1}}}\right|\left|\vec{k_{2}}\right|}\left(\left|\vec{k}_{1}\right|+\left|\vec{k}_{2}\right|\right)\right.\\
\left.-\frac{\left(\left|\vec{k}_{2}\right|^{2}-\left|\vec{k}_{1}\right|^{2}\right)\left(\left|\vec{k}_{2}\right|-\left|\vec{k}_{1}\right|\right)}{\left|\vec{k_{_{1}}}\right|\left|\vec{k_{2}}\right|}\right]\\
=-\frac{i}{2}\frac{1}{\left(E_{p}^{2}-\left(\left|\vec{k}_{1}\right|-\left|\vec{k}_{2}\right|\right)^{2}\right)\left(E_{p}^{2}-\left(\left|\vec{k}_{1}\right|+\left|\vec{k}_{2}\right|\right)^{2}\right)}\frac{\left(\left|\vec{k}_{1}\right|+\left|\vec{k}_{2}\right|\right)}{\left|\vec{k_{_{1}}}\right|\left|\vec{k_{2}}\right|}\left[E_{p}^{2}-\left(\left|\vec{k}_{2}\right|-\left|\vec{k}_{1}\right|\right)^{2}\right]\\
=-\frac{i}{2}\frac{1}{E_{p}^{2}-\left(\left|\vec{k}_{1}\right|+\left|\vec{k}_{2}\right|\right)^{2}}\left(\frac{1}{\left|\vec{k_{_{1}}}\right|}+\frac{1}{\left|\vec{k_{_{2}}}\right|}\right)
\end{multline}
which is what we set out to prove in the beginning.

\section{Acknowledgments }

The origins of this project go back some time. JPR is grateful to Antal Jevicki and Robert de Mello Koch for their early interest in the project, and for insightful comments, particularly Robert de Mello Koch, on a recent draft of this paper. 

\pagebreak{}

\end{document}